\documentclass[%
 aip,
 amsmath,amssymb,
 reprint,%
]{revtex4-1}
\usepackage{lineno}
\usepackage{graphicx}
\usepackage{dcolumn}
\usepackage{bm}

\usepackage[utf8]{inputenc}
\usepackage[T1]{fontenc}
\usepackage{mathptmx}

\begin{document}

\preprint{AIP/123-QED}

\title[]{Spontaneous imbibition dynamics in interacting multi-capillary systems: A generalized model}

\author{Shabina Ashraf}
\affiliation{Indian Institute of Technology Delhi, Hauz Khas, New Delhi, India. 110016.}
\affiliation{University of Rennes1, Rennes, France. 35708.}
\author{Yves M{\'e}heust}
\affiliation{University of Rennes1, Rennes, France. 35708.}
\author{Jyoti Phirani}
\email{jphirani@chemical.iitd.ac.in}
\affiliation{Indian Institute of Technology Delhi, Hauz Khas, New Delhi, India. 110016.}



\date{\today}

\begin{abstract}
Bundle-of-tubes model was previously used to understand the flow behaviour in a porous medium. The interacting nature of the pores within a porous medium can be well depicted by an interacting capillary model. However, the arrangement of pores is crucial in understanding the flow behaviour in an interacting capillary system, which also leads to different governing equations of spontaneous imbibition. To this end, in the present work, we first develop a generalized one-dimensional lubrication approximation model to predict the imbibition behaviour in an interacting multi-capillary system. Using our generalized model, we observe that the flow dynamics, the capillary having the leading meniscus and the breakthrough time are governed by the contrast in the radii and the arrangement of the capillaries. We also show that during breakthrough, the saturation of the multi-capillary system depends on the arrangement of the capillaries. We show that the breakthrough in the bundle-of-tubes model occurs at a dimensionless time of $0.5$, while the breakthrough in the interacting capillary system occurs between the dimensionless times $0.31$ and $0.42$, for the capillary system considered in this study. Comparing the interacting multi-capillary system with the bundle-of-tubes model, we present substantial deviations and show that the interacting capillary system is closer to the real porous medium.

\end{abstract}

\maketitle
\section{Introduction}

When a wetting fluid is placed in contact with a porous medium, the fluid spontaneously imbibes into the pore spaces due to capillary suction. The imbibition of the wetting fluid in the porous matrix is crucial for oil recovery from reservoirs\citep{xiao2012prediction,lin2017characterizing,saraji2010adsorption}, Paper Analytic Devices ($\mu$PADs)\citep{taghizadeh2019paper,soda2019equipment}, textiles\citep{dai2019bioinspired}, inkjet printing\citep{rosello2019dripping,wang2019fabrication}, microfluidics\citep{liu2019microfluidic,gharibshahi2020hybridization,carrell2019,schaumburg}, lab-on-chip devices\citep{lin2008performance,lee2019whole,joung2019based}, diagnostics\citep{liang2019,rich}, Polymer Electrolyte Membrane Fuel Cell (PEMFC)\citep{xiao2019novel,carrere2019liquid}, micro heat pipes\citep{singh2018enhanced,chernysheva2019simulation}, in understanding the motion of blood cells\citep{pozrikidis2005axisymmetric} and in the design of bio-inspired drainage and ventilation systems\citep{singh2019architectural}. The capillary driven imbibition in a homogeneous porous medium follows diffusive dynamics, where the length of imbibition is proportional to the square root of time\citep{li2015criteria,gruener2019capillarity}. This kind of diffusive dynamics for spontaneous imbibition was first seen by Lucas\citep{Lucas} and Washburn\citep{Washburn} for a cylindrical capillary tube. They report that, the spontaneous imbibition of a wetting fluid of viscosity $\mu$ in a horizontal capillary tube of radius $r$, is given by,
\begin{equation}
    \label{eq02}
    l = \sqrt{\frac{r\sigma \cos{\theta}}{2\mu}t},
\end{equation}
where, $l$ is the length advanced by the meniscus in time $t$, $\sigma$ is the surface tension and $\theta$ is the contact angle made by the invading liquid with the capillary wall. Eq. \ref{eq02} shows that, the length advanced by the invading fluid is proportional to the square root of time $t$ and the radius of the capillary $r$. This implies that, in time $t$, the meniscus will advance more in a larger radius capillary as compared to a capillary of smaller radius. 
Later, the phenomenon of imbibition in a single pore has been observed to be strongly dependent on the geometries of the capillaries\citep{lenormand1984role,dong1995imbibition,ramezanzadeh2019simulating,zheng2019integrated,reyssat2008,budaraju2016capillary,ouali2013wetting,rosendahl2010convective,weislogel2001capillary,dimitrov2008flow}. 

Due to the similarity of imbibition phenomenon in a capillary tube and homogeneus porous medium, a bundle of non-interacting capillaries have been considered as a proxy porous medium\citep{dahle2005bundle,douglas2011capillary,bartley2001relative,bartley1999relative}. However, in a naturally occurring porous medium, the pores are of various shapes, sizes and are interconnected. To this end, several studies explored the fluid flow in interacting capillaries to understand the effect of interaction on the pore scale flow dynamics\citep{Ashraf2018,Ashraf2017,dong2005immiscible,dong2006immiscible,wang2008fluid,dong1998characterization,li2017crossflow,krishnamurthy2007gas}. In an interacting two capillary system, the imbibition in the small radius capillary is found to be faster unlike the diffusive dynamics given by Lucas and Washburn shown in Eq. \ref{eq02}. Unsal et al.,\citep{unsal2007,unsal2007co,unsal2009bubble} experimentally showed that, the imbibition speed is fastest in the capillary having the least effective radius, for a non-cylindrical interacting three capillary system. On the contrary, Ashraf et al.,\citep{Ashraf2018} showed that, the imbibition is not always fastest in the smallest radius capillary in a cylindrical interacting three capillary system using a one-dimensional lubrication approximation model.

If we want to use a tube bundle model to fundamentally understand the flow behavior in a porous medium, we need to use interacting capillaries. To that end, in this work, we develop a generalized one-dimensional model to predict the spontaneous imbibition in the randomly arranged capillaries that interact with each other. We review the work done by Ashraf et al.,\citep{Ashraf2018} for two and three interacting capillary systems. Then, we proceed to understand the imbibition dynamics in an interacting four-capillary system. We explain the underlying physical phenomena causing the menisci advance at different rates in each of the capillaries. We show that, the contrast in the capillary radii and their arrangement in the interacting capillary system is crucial in predicting the imbibition dynamics. We propose a generalized one-dimensional model to predict the imbibition in an interacting multi-capillary system. We show that, by varying one or both the influencing parameters, namely the contrast in radii and the arrangement of capillaries in the interacting multi-capillary system, the imbibition dynamics change significantly.

\section{Capillary imbibition in interacting capillaries}

\begin{figure}[h!]
    \centering
    \includegraphics[width=6 cm]{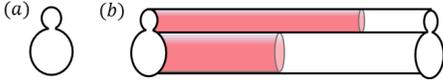}
    \caption{Spontaneous imbibition in two interacting capillaries, (a) cross section view, (b) longitudinal view}
    \label{f1}
\end{figure}

Using capillary system shown in Fig. \ref{f1}, Ashraf et al.,\citep{Ashraf2018} used VOF simulations to develop a reduced order, Washburn-like one dimensional model for two and three interacting capillaries. The capillaries can interact hydrodynamically with the neighbouring capillaries along the axes. Their model predicts that, the meniscus in the smaller radius capillary leads during the spontaneous imbibition in a two-interacting capillary system, as shown by their VOF simulations. This is because, (1) the pressure is same in the sections of the capillary system where both the capillaries are filled with the same fluid; and (2) the invading fluid transfers from the large radius capillary to the small radius capillary at the meniscus of the large radius capillary\citep{Ashraf2018}. Their results show that, if a bundle-of-tubes model is to be used to understand the flow behaviour in a real porous medium like an oil reservoir, it is essential to consider the interaction between the tubes.

For an interacting three-capillary system, the model of Ashraf et al.,\citep{Ashraf2018} showed that, the radii and the arrangement of the capillaries significantly change the imbibition behavior in the capillary system and the meniscus in the small radius capillary does not always lead. To this end, we develop a generalized multiple interacting capillaries model for spontaneous imbibition. We first describe the one-dimensional model formulation for an interacting four-capillary system to understand the underlying equations for generalization.

\section{Model development for four interacting capillaries}
\label{model_dev}
To find the flow properties of a porous medium using a system of interacting capillaries, we need to take into account, the arrangement of capillaries, unlike in the bundle-of-tubes model. For a porous medium made of $n$ interacting capillaries, we can have $\frac{n!}{2}$ arrangements. 
\begin{figure}[h!]
    \centering
    \includegraphics[width=8 cm]{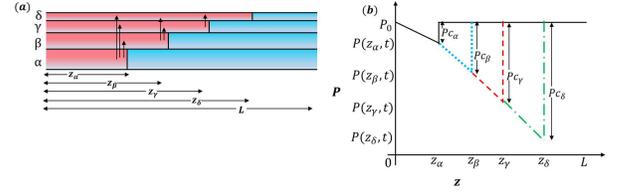}
    \caption{Spontaneous imbibition in an ordered interacting four-capillary system. (a) Schematic showing the imbibition phenomenon. The imbibition lengths in capillaries $\alpha$, $\beta$, $\gamma$, $\delta$ of radii $r_\alpha$, $r_\beta$, $r_\gamma$, $r_\delta$ are denoted by $z_\alpha$, $z_\beta$, $z_\gamma$, $z_\delta$, respectively. (b) The pressure vs imbibition distance plot for an ordered interacting four-capillary system showing that the imbibition in an ordered interacting capillary system is $z_\alpha<z_\beta<z_\gamma<z_\delta$.}
    \label{f2}
\end{figure}
Fig. \ref{f2}(a) shows an ordered arrangement of four capillaries. The radii of the capillaries are $r_\alpha$, $r_\beta$, $r_\gamma$, $r_\delta$, such that $r_\alpha > r_\beta > r_\gamma > r_\delta$ and we call the arrangement $\alpha\beta\gamma\delta$. The capillary pressure for the invading fluid, making a contact angle $\theta_w$ with the capillary wall and having a surface tension $\sigma$, is given by Young-Laplace equation as $Pc_i=\frac{2\sigma cos\theta_w}{r_i}(i=\alpha,\beta,\gamma\delta)$\citep{young,laplace}, therefore $Pc_\alpha<Pc_\beta<Pc_\gamma<Pc_\delta$. We consider the assumptions from Ashraf et al.,\citep{Ashraf2018} that, (1) the pressure equilibrates in the sections of the capillary system filled with the invading fluid and (2) fluid transfers from a capillary having a larger radius to the capillary having a smaller radius just before the meniscus, which is assumed to occur `at' the meniscus. We also consider the interaction between the capillaries to be sufficiently low such that, the poiseuille flow in each of the capillaries is maintained. We show the direction of fluid transfer in the interacting capillary system using arrows in the schematic Fig. \ref{f2}(a). At a given time $t$, from the inlet till the first fluid front, say $z_\alpha$, the pressure drop in all the capillaries is same as  shown in Fig. \ref{f2}(b). The pressure jump across the first fluid front will be equal to the capillary pressure of the capillary having the smallest imbibition length. We know that, the capillary pressure is least i.e., $Pc_\alpha$ in the capillary $\alpha$, therefore, the imbibed length is $z_\alpha$ in the capillary having radius $r_\alpha$. At this time $t$, the pressure in capillaries $\beta,\gamma,\delta$ is same after $z_\alpha$ till the meniscus in the capillary $\beta$ located at $z_\beta$, as shown by the dotted line in Fig. \ref{f2}(b). At $z_\beta$, the capillary pressure jump is $Pc_\beta$. From $z_\beta$ to the imbibition length in capillary $\gamma$, i.e., $z_\gamma$, the pressure gradient is same in capillaries $\gamma,\delta$ as shown by the dashed line in Fig. \ref{f2}(b). The imbibition length in capillary $\delta$ leads the other three capillaries shown by the dash-dot line in Fig. \ref{f2}(b). We can see that, Fig. \ref{f2}(b) shows the only possible pressure gradient profile and imbibition behaviour for the ordered interacting four-capillary system. 

\begin{figure}[h!]
    \centering
    \includegraphics[width=8 cm]{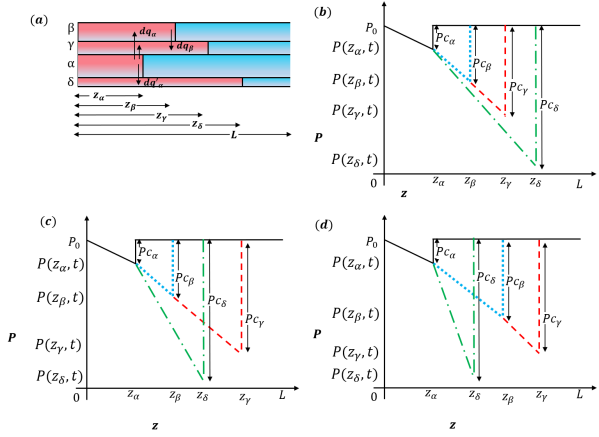}
    \caption{Spontaneous imbibition in an interacting four-capillary system where the order of arrangement of capillaries is $\beta\gamma\alpha\delta$. (a) Schematic of the imbibition phenomenon showing the fluid transfer at menisci locations. The imbibition lengths in capillaries $\alpha$, $\beta$, $\gamma$, $\delta$ of radii $r_\alpha$, $r_\beta$, $r_\gamma$, $r_\delta$ are denoted by $z_\alpha$, $z_\beta$, $z_\gamma$, $z_\delta$, respectively. (b), (c), (d) show the possible pressure vs imbibition distance plot for the interacting four-capillary system shown in (a). Depending upon the contrast in radii of the capillaries, the possible cases for the pressure vs distance plots are (b) where $z_\alpha<z_\beta<z_\gamma<z_\delta$, (c) where $z_\alpha<z_\beta<z_\delta<z_\gamma$ and (d) where $z_\alpha<z_\delta<z_\beta<z_\gamma$.}
    \label{f3}
\end{figure}

We now consider one of the random arrangements as shown in the schematic of Fig. \ref{f3}(a), where the order of arrangement of the capillaries is $\beta\gamma\alpha\delta$. This random arrangement is elaborated to describe the model development for predicting the imbibition phenomenon in a randomly arranged interacting capillary system. For this arrangement, depending upon the contrast in radii, three different pressure gradient plots are possible as shown in Fig. \ref{f3}(b),(c),(d). At a given time $t$, from the inlet till the first meniscus at $z_\alpha$, the pressure gradient in all the capillaries is same as shown in all the possible pressure gradient plots of Fig. \ref{f3}(b),(c),(d). After $z_\alpha$, the imbibing fluid is continuous in the capillaries $\beta,\gamma$ and therefore, the pressure gradient is same in these capillaries after the first meniscus, till $z_\beta$. Although the capillary $\delta$ is filled with the imbibing phase, the non-wetting fluid in $\alpha$ disconnects it from capillaries $\beta, \gamma$. Therefore, after $z_\alpha$, the pressure gradient in $\delta$ can be different from the pressure gradient in $\beta,\gamma$, as shown in the pressure gradient plots. As the capillary suction of $\beta$ is less than the capillary suction in $\gamma$, the meniscus in $\beta$ precedes the meniscus in $\gamma$, implying, $z_\beta<z_\gamma$, at all times. Therefore, for the arrangement $\beta\gamma\alpha\delta$ shown in the schematic of Fig. \ref{f3}(a), $z_\alpha<z_\beta<z_\gamma$ and $z_\alpha<z_\delta$, during the imbibition phenomenon. The position of $z_\delta$ relative to $z_\beta$ and $z_\gamma$ depends on the contrast of capillaries' radii.
We now proceed to develop a one-dimensional model to predict the imbibition behaviour in the interacting four-capillary system $\beta\gamma\alpha\delta$ as shown in the schematic of Fig. \ref{f3}(a). From the inlet till the first meniscus in capillary $\alpha$, the pressure drop can be given by Hagen-Poiseuille's law as,
\begin{equation}
    \label{eq03}
    P(z_\alpha,t) - P_0 = -\frac{8\mu z_\alpha}{r_\alpha^2} v_\alpha(t),
\end{equation}
where, $\mu$ is the imbibing fluid viscosity, $v_\alpha(t)$ is the instantaneous velocity of the wetting fluid in the capillary $\alpha$ from the inlet till $z_\alpha$, $P_0$ is the inlet pressure and $P(z_\alpha,t)$ is the pressure in the imbibing fluid at $z_\alpha$, as shown in Fig. \ref{f3}. Till the first fluid front at $z_\alpha$, the pressure gradient in all the capillaries is same as shown in Fig. \ref{f3}(b),(c),(d), which by Poiseuille's equation implies,
\begin{equation}
    \label{eq04}
    \frac{v_\alpha(t)}{r_\alpha^2} = \frac{v_\beta(t)}{r_\beta^2} = \frac{v_\gamma(t)}{r_\gamma^2} = \frac{v_\delta(t)}{r_\delta^2},
\end{equation}
where, $r_i$ $(i=\alpha, \beta, \gamma, \delta)$ is the radius of the capillary and $v_i(t)$ $(i=\alpha, \beta, \gamma, \delta)$ is the velocity from inlet till $z_\alpha$ in the capillary having radius $r_i$. The capillary pressure jump is $Pc_\alpha$ at $z_\alpha$ and some of the impregnating fluid transfers from the capillary $\alpha$ to other capillaries. The volumetric fluid transfer from $\alpha$ to $\beta,\gamma$ is, say, $dq_\alpha$ and from $\alpha$ to $\delta$ is $dq'_\alpha$. The velocity of the advancing meniscus in $\alpha$ is $\frac{dz_\alpha}{dt}$ given by,
\begin{equation}
    \label{eq05}
    \frac{dz_\alpha}{dt} = v_\alpha(t)-\frac{(dq_\alpha+dq'_\alpha)}{\pi r_\alpha^2},
\end{equation}
From $z_\alpha$ till $z_\delta$, the velocity of the fluid in the capillary $\delta$, given by $\frac{dz_\delta}{dt}$ is,
\begin{equation}
    \label{eq06}
    \frac{dz_\delta}{dt} = v_\delta(t)+\frac{dq'_\alpha}{\pi r_\delta^2},
\end{equation}
So, the pressure drop in the capillary $\delta$ from $z_\alpha$ till $z_\delta$ is,
\begin{equation}
    \label{eq07}
    P(z_\delta,t) - P(z_\alpha,t) = -\frac{8\mu (z_\delta-z_\alpha)}{r_\delta^2} \left(v_\delta(t)+\frac{dq'_\alpha}{\pi r_\delta^2}\right),
\end{equation}
where, $P(z_\delta,t)$ is the pressure of the impregnating fluid at $z_\delta$ as shown in Fig. \ref{f3}. At $z_\delta$, the pressure jump across the meniscus is $Pc_\delta$, as shown in the pressure gradient plots of Fig. \ref{f3}(b),(c),(d).

The capillaries $\beta$ and $\gamma$ are on the other side of the capillary $\alpha$ and the imbibing fluid in these capillaries is continuous after $z_\alpha$ as shown in the schematic of Fig. \ref{f3}(a). Therefore, the pressure gradient in the capillaries $\beta$, $\gamma$ is same after $z_\alpha$, which gives the condition,
\begin{equation}
    \label{eq08}
    \frac{v_\beta(t)+\frac{\omega dq_\alpha}{\pi r_\beta^2}}{\pi r_\beta^2} = \frac{v_\gamma(t)+\frac{(1-\omega)dq_\alpha}{\pi r_\gamma^2}}{\pi r_\gamma^2},
\end{equation}
where $\omega$ and $(1-\omega)$ are the fractions of volumetric fluid transfer $dq_\alpha$ to the capillaries $\beta$ and $\gamma$, respectively. From Eq. \ref{eq04} and Eq. \ref{eq08}, we obtain the fraction $\omega=\frac{r_\beta^4}{r_\beta^4+r_\gamma^4}$. As the capillary pressure jump of the capillary $\beta$ is less than the capillary pressure jump in the capillary $\gamma$, the meniscus in $\beta$ lags behind the meniscus in $\gamma$. Therefore, the pressure drop in capillaries $\beta$ and $\gamma$ from $z_\alpha$ to $z_\beta$ is,
\begin{equation}
    \label{eq09}
    P(z_\beta,t) - P(z_\alpha,t) = -\frac{8\mu (z_\beta-z_\alpha)}{r_\beta^2} \left(v_\beta(t)+\omega\frac{dq_\alpha}{A_\beta}\right),
\end{equation}
where, $P(z_\beta,t)$ is the pressure in the imbibing fluid at $z_\beta$. At the meniscus in capillary $\beta$, the capillary pressure jump is $Pc_\beta$ and some of the impregnating fluid transfers from $\beta$ to $\gamma$, which we assume to be $dq_\beta$. The velocity of meniscus in $\beta$ after the fluid transfer $dq_\beta$ is,
\begin{equation}
    \label{eq10}
    \frac{dz_\beta}{dt} = v_\beta(t)+\omega\frac{dq_\alpha}{\pi r_\beta^2}-\frac{dq_\beta}{\pi r_\beta^2},
\end{equation}
After $z_\beta$, the meniscus in the capillary $\gamma$ travels with a velocity given by,
\begin{equation}
    \label{eq11}
    \frac{dz_\gamma}{dt} = v_\gamma(t)+(1-\omega)\frac{dq_\alpha}{\pi r_\gamma^2}+\frac{dq_\beta}{\pi r_\gamma^2}.
\end{equation}
The pressure drop from $z_\beta$ to $z_\gamma$ in capillary $\gamma$ is given by,
\begin{equation}
    \label{eq12}
    P(z_\gamma,t) - P(z_\beta,t) = -\frac{8\mu (z_\gamma-z_\beta)}{r_\gamma^2} \left(v_\gamma(t)+(1-\omega)\frac{dq_\alpha}{\pi r_\gamma^2}+\frac{dq_\beta}{\pi r_\gamma^2}\right),
\end{equation}
where, $P(z_\gamma,t)$ is the pressure in the impregnating fluid at $z_\gamma$. The capillary pressure jump across the meniscus in the capillary $\gamma$ is $Pc_\gamma$. The pressure jump across the menisci in each of the capillaries in the interacting four-capillary system is given by the Young-Laplace equation\citep{young,laplace}, that is,
\begin{equation}
  \label{eq13}
   P(z_i,t)-P_0 = -Pc_i = -\frac{c\sigma \cos\theta_w}{r_i},
\end{equation}
where, $i=\alpha, \beta, \gamma, \delta$. In Eq.\ref{eq13}, $\sigma$ is the interfacial tension  and $\theta_w$ is the contact angle of the invading fluid with the capillary surface, the constant $c$ depends on the geometry of the capillaries and $c=2$ for a cylindrical cross-section geometry. Eq. \ref{eq13} gives us the pressure drop in  each of the sections and by substituting Eqs. \ref{eq05}, \ref{eq06}, \ref{eq10}, \ref{eq11} in Eqs. \ref{eq03}, \ref{eq07}, \ref{eq09}, \ref{eq12}, the equations governing the flow in the interacting capillary system are given as,
\begin{equation}
  \label{eq14}
   Pc_\alpha = \frac{8\mu z_\alpha}{r_\alpha^4+r_\beta^4+r_\gamma^4+r_\delta^4} \left(r_\alpha^2\frac{dz_\alpha}{dt}+r_\beta^2\frac{dz_\beta}{dt}+r_\gamma^2\frac{dz_\gamma}{dt}+r_\delta^2\frac{dz_\delta}{dt} \right),
\end{equation}
\begin{equation}
  \label{eq15}
   Pc_\delta - Pc_\alpha = \frac{8\mu (z_\delta-z_\alpha)}{r_\delta^2} \left(\frac{dz_\delta}{dt} \right),
\end{equation}
\begin{equation}
  \label{eq16}
   Pc_\beta - Pc_\alpha = \frac{8\mu (z_\beta-z_\alpha)}{r_\beta^4+r_\gamma^4} \left(r_\beta^2\frac{dz_\beta}{dt}+r_\gamma^2\frac{dz_\gamma}{dt}\right),
\end{equation}
\begin{equation}
  \label{eq17}
   Pc_\gamma - Pc_\beta = \frac{8\mu (z_\gamma-z_\beta)}{r_\gamma^2} \left(\frac{dz_\gamma}{dt} \right).
\end{equation}

The Eqs. \ref{eq14} to \ref{eq17} are made dimensionless by considering dimensionless lengths as $Z_i=\frac{z_i}{L}$ where $i=\alpha, \beta, \gamma, \delta$ and $L$ is the total length of the capillary system. The time is non-dimensionalised as $T=\frac{Pc_\alpha r_\alpha^2}{8\mu L^2}t$. The dimensionless form of Eqs. \ref{eq14} to \ref{eq17} are,
\begin{equation}
  \label{eq18}
   1 = \frac{Z_\alpha}{1+\lambda_\beta^4+\lambda_\gamma^4+\lambda_\delta^4}\left(\frac{dZ_\alpha}{dT}+\lambda_\beta^2\frac{dZ_\beta}{dT}+\lambda_\gamma^2\frac{dZ_\gamma}{dT}+\lambda_\delta^2\frac{dZ_\delta}{dT} \right),
\end{equation}
\begin{equation}
  \label{eq19}
   \epsilon_\delta - 1 = \frac{Z_\delta-Z_\alpha}{\lambda_\delta^2} \left(\frac{dZ_\delta}{dT} \right),
\end{equation}
\begin{equation}
  \label{eq20}
   \epsilon_\beta - 1 = \frac{Z_\beta-Z_\alpha}{\lambda_\beta^4+\lambda_\gamma^4} \left(\lambda_\beta^2\frac{dZ_\beta}{dT}+\lambda_\gamma^2\frac{dZ_\gamma}{dT}\right),
\end{equation}
\begin{equation}
  \label{eq21}
   \epsilon_\gamma - \epsilon_\beta = \frac{Z_\gamma-Z_\beta}{\lambda_\gamma^2} \left(\frac{dZ_\gamma}{dT} \right),
\end{equation}
where, $\lambda_i=\frac{r_i}{r_\alpha}$ and $\epsilon_i=\frac{Pc_i}{Pc_\alpha}$ for $i=\beta, \gamma, \delta$. Further, if the contact angle $\theta_w$ is same for all the capillaries, $\epsilon_i=\frac{1}{\lambda_i}$.
By rearranging the governing Eqs. \ref{eq18} to \ref{eq21} and adding them gives,
\begin{equation}
    \label{eq22}
    2\left(1+\sum_{i=\beta,\gamma,\delta} \epsilon_i \lambda_i^4\right)T = Z_\alpha^2+Z_\beta^2 \lambda_\beta^2+Z_\gamma^2 \lambda_\gamma^2+Z_\delta^2 \lambda_\delta^2.
\end{equation}
 Eq. \ref{eq22} describes that, in an interacting capillary system, the sum of the squares of the product of dimensionless radius with the dimensionless distance invaded in all the capillaries is proportional to the invasion time $T$. For different arrangements of an interacting four-capillary system having the same contrast in capillary radii, the total capillary suction of the system remains the same. Therefore, for all the $\frac{4!}{2}=12$ arrangements of an interacting four-capillary system, rearranging the equations governing the imbibition and adding them gives Eq. \ref{eq22}. Therefore, for a system having $n$-interacting capillaries, the imbibition phenomenon can be described as,
\begin{equation}
    \label{eq23}
    2\left(1+\sum_{i=1}^n \epsilon_i \lambda_i^4\right)T = \sum_{i=1}^n \psi_j Z_j
\end{equation}
where $\psi_i=\frac{\pi r_i^2 z_i}{\pi r_\alpha^2 L}$ $(j=1,2,...,n)$ are the dimensionless volume imbibed imbibed in each of the capillaries. Eq. \ref{eq23} shows that the sum of dimensionless volume times the dimensionless length invaded is proportional to time. We know that the imbibition in non-interacting capillaries follows diffusive dynamics, where the product of volume times length is proportional to time, whereas for interacting capillaries, the sum of the volume times the invaded length in all the capillaries is proportional to time. We note from the above derivations that each arrangement will have a different set of governing equations for menisci positions with time. This is because, the knowledge of the sequence of the menisci is required for applying the assumptions of `pressure equilibration' and `fluid transfer'. Therefore, for an interacting $n$-capillary system, we now develop an algorithm which can determine the imbibition behaviour in the interacting capillary system and form the governing equations.

\section{Generalizing the one-dimensional spontaneous imbibition model in the interacting capillary system}
At time $t$, the meniscus in the largest radius capillary lags and divides the capillaries into two regions. 
\begin{figure}[h!]
    \centering
    \includegraphics[width=8 cm]{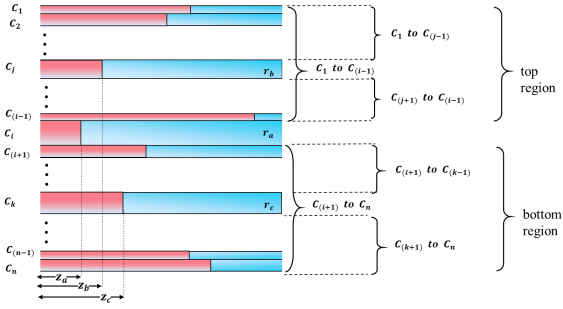}
    \caption{Schematic of spontaneous imbibition in an $n$-capillary system where the capillaries are positioned randomly. The capillaries in the sequence of arrangement are denoted by $C_1, C_2, ... C_n$. The capillary radii are denoted as $r_a, r_b, r_c,...$ and the corresponding imbibition distances are denoted by $z_a, z_b, z_c,...$.}
    \label{f4}
\end{figure}
The schematic of a multiple interacting capillary system is shown in Fig. \ref{f4}. The capillaries in the order of arrangement are numbered from $C_1$ to $C_n$. The largest radius capillary has a radius $r_a$, positioned at $C_i$ and has a capillary pressure $Pc_a$, as shown in Fig. \ref{f4}. We observe from the four-interacting capillary system that, we first need to find the largest radius capillary. The capillary $C_i$ divides the capillary system into two regions, that is the top region and the bottom region, as shown in Fig. \ref{f4}. In the top region, the largest radius capillary positioned at $C_j$, has a radius $r_b$ and the corresponding capillary pressure is $Pc_b$. Similarly, in the bottom region, the largest capillary has a radius $r_c$ and a capillary pressure $Pc_c$, which is positioned at $C_k$. In developing a generalized model for the interacting capillary system shown in Fig. \ref{f4}, the following step-by-step procedure is followed.
\begin{enumerate}
\item{We initiate the model formulation by choosing the capillary with the largest radius, i.e., $r_a$ as shown in Fig. \ref{f4}. The pressure gradient in all the capillaries is same till $z_a$, which is determined by Hagen-Poiseuille's law. Some of the invading fluid from the largest radius capillary transfers to other capillaries just before the meniscus and is considered to occur `at' $z_a$.}
    \item{We see from Fig. \ref{f4} that, the imbibing fluid in capillaries $C_1$ to $C_{(i-1)}$ is separated from the imbibing fluid in the capillaries $C_{(i+1)}$ to $C_n$. Therefore, on one side of capillary $r_a$, the capillaries $C_1$ to $C_{(i-1)}$ are classified into one region, while the capillaries from $C_{(i+1)}$ to $C_n$ on the other side of $r_a$ are classified into another region. The fluid transfer from the capillary of radius $r_a$ is divided among the other capillaries according to their radii. The fluid transfer from $C_i$ causes the flow rate to increase in capillaries $C_1$ to $C_{(i-1)}$ and $C_{(i+1)}$ to $C_n$.}
    \item{The large radius capillary among the capillaries $C_1$ to $C_{(i-1)}$ is identified, which is capillary $r_b$, located at $C_j$, as shown in the schematic of Fig. \ref{f4}. The pressure from $z_a$ to $z_b$ is same in capillaries $C_1$ to $C_{(i-1)}$, which is determined by Hagen-Poiseuille's law. Again, `at' $z_b$, some of the invading fluid transfers from $C_j$ to the capillaries $C_1$ to $C_{(j-1)}$ and $C_{(j+1)}$ to $C_{(i-1)}$, which increases the flow rate in these capillaries.}
    \item{Similarly, the largest radius capillary from $C_{(i+1)}$ to $C_n$ is chosen, which is capillary $r_c$, located at $k$, as shown in Fig. \ref{f4}. The pressure gradient is same in the capillaries $C_{(i+1)}$ to $C_n$ from $z_a$ till $z_c$, which is determined. Then, `at' $z_c$, some of the invading fluid transfers into the capillaries $C_{(i+1)}$ to $C_{(k-1)}$ and $C_{(k+1)}$ to $C_n$, which causes an increase in flow rate in these capillaries.}
    \item{We see from Fig. \ref{f4} that, the impregnating fluid in the regions $C_1$ to $C_{(j-1)}$ and $C_{(j+1)}$ to $C_{(i-1)}$ are separated by capillary $C_j$. Again, the large radius capillary among the capillaries $C_1$ to $C_{(j-1)}$ is identified. Similarly, the large radius capillary among the capillaries $C_{(j+1)}$ to $C_{(i-1)}$ is also identified. The pressure in each of these regions are determined by Hagen-Poiseuille's law and the fluid transfers from the largest radius capillary of the respective regions are also considered.}
    \item{In the regions $C_{(i+1)}$ to $C_{(k-1)}$ and $C_{(k+1)}$ to $C_n$, the large radius capillaries are chosen and the respective pressure drop till the meniscus in the large radius capillaries in these regions are determined. The fluid transfer in the newly formed regions are also considered to determine pressure drop in each of the regions.}
    \item{The division of capillaries into regions is repeated until there is only one capillary in a region, whose pressure drop is determined. For an interacting capillary system consisting of $n$ capillaries, the capillaries are divided in $n$ regions.}
    \item{The pressure jump across the meniscus in each of the capillaries is the corresponding Young-Laplace capillary pressure of that capillary. The $n$ pressure drop equations are then solved to obtain the lengths impregnated in each of the capillaries.}
\end{enumerate}
By following the step wise procedure, the equations governing the imbibition in the capillaries are formulated. They are non-dimensionalized as explained. A MATLAB program is used to solve the governing equations to obtain the advancement of menisci in the capillaries with time.

\section{Results and discussions}

We first explore the imbibition behavior in an interacting four-capillary system followed by the imbibition in an interacting multi-capillary system consisting of $20$ capillaries. 
\subsection{Interacting four-capillary system}

\begin{figure}[h!]
    \centering
    \includegraphics[width=8 cm]{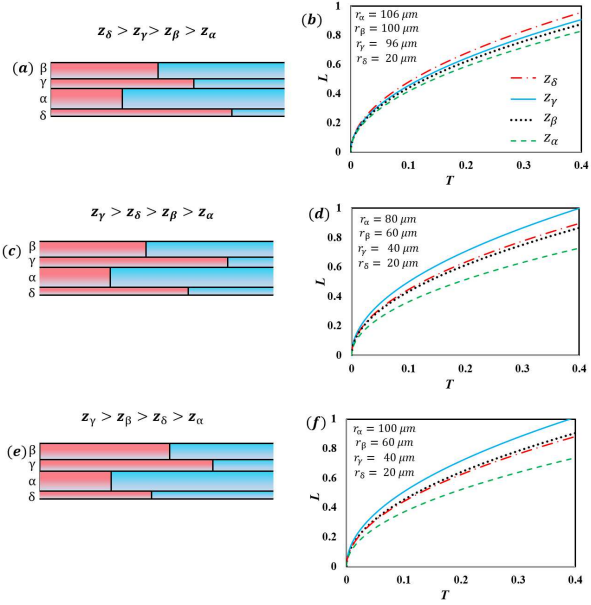}
    \caption{Spontaneous imbibition in interacting four-capillary system where the capillaries are placed in the arrangement $\beta\gamma\alpha\delta$. (a), (c), (e) represent the schematics of possible imbibition behavior. The dimensionless length vs dimensionless time graphs in (b), (d), (f) show the corresponding contrast in radii for which the imbibition behavior represented in (a), (c), (e) are observed. The time at which the leading meniscus reaches the outlet end of the interacting capillary system $(T_{bt})$ for the cases (b), (d) and (f) are $0.43$, $0.39$ and $0.39$, respectively.}
    \label{f5}
\end{figure}
In section \ref{model_dev}, using pressure vs length graphs, we anticipate that, in an ordered arrangement, the meniscus in the smallest radius capillary $\delta$ will always lead, followed by the second smallest radius capillary $\gamma$ as shown in Fig. \ref{f2}, while the meniscus in the capillary $\alpha$ always lags. Solving the governing equations for this arrangement, we always get the same trend of the imbibed lengths in the capillaries at any given time during the imbibition process. However, $\frac{4!}{2}=12$ arrangements are possible for an interacting four-capillary system, for any given $4$ radii of the capillaries. We chose one arrangement named $\beta\alpha\gamma\delta$ in section \ref{model_dev} and anticipated $3$ cases based on the pressure gradient plots as shown in Fig. \ref{f3}(b), (c), (d). 

Solving Eqs. \ref{eq18} to \ref{eq21}, we show in Fig. \ref{f5} that all the three pressure vs length graphs shown in Fig. \ref{f3} are possible by changing the contrast in the radii of capillaries. The relative positions of $Z_\beta$, $Z_\gamma$, $Z_\delta$ change due to the contrast of radii in the interacting capillary system. 
\begin{figure}
    \centering
    \includegraphics[width=8 cm]{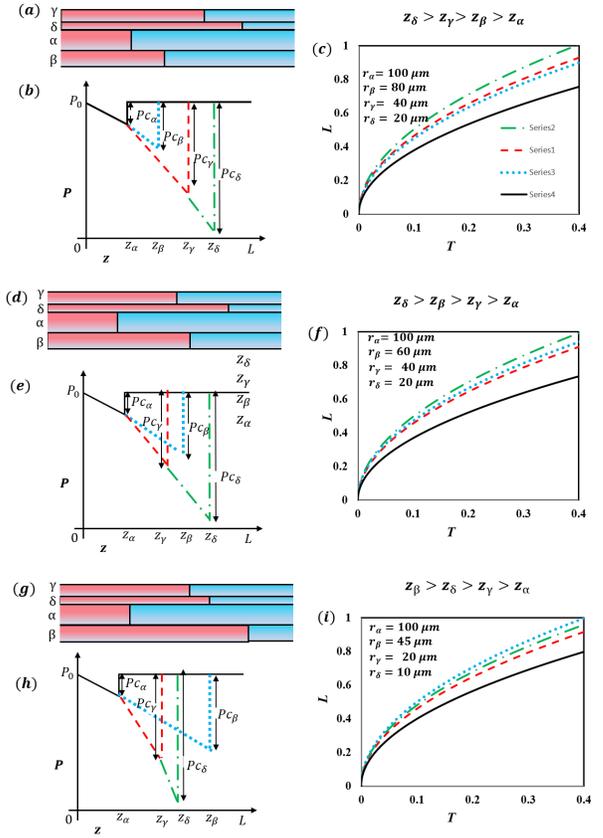}
    \caption{Spontaneous imbibition in interacting four-capillary system where the capillaries are placed in the arrangement $\gamma\delta\alpha\beta$. (a), (d), (g) represent the schematics of possible imbibition behavior, (b), (e), (h) show the corresponding pressure gradient plots. The dimensionless length vs dimensionless time graphs in (c), (f), (i) show radii for which the imbibition behavior shown in (a), (d), (g) is observed. The time at which the invading fluid reaches the outlet end $(T_{bt})$ for the cases (c), (f) and (i) are $0.38$, $0.40$ and $0.39$, respectively}
    \label{f6}
\end{figure}
\begin{figure}
    \centering
    \includegraphics[width=8 cm]{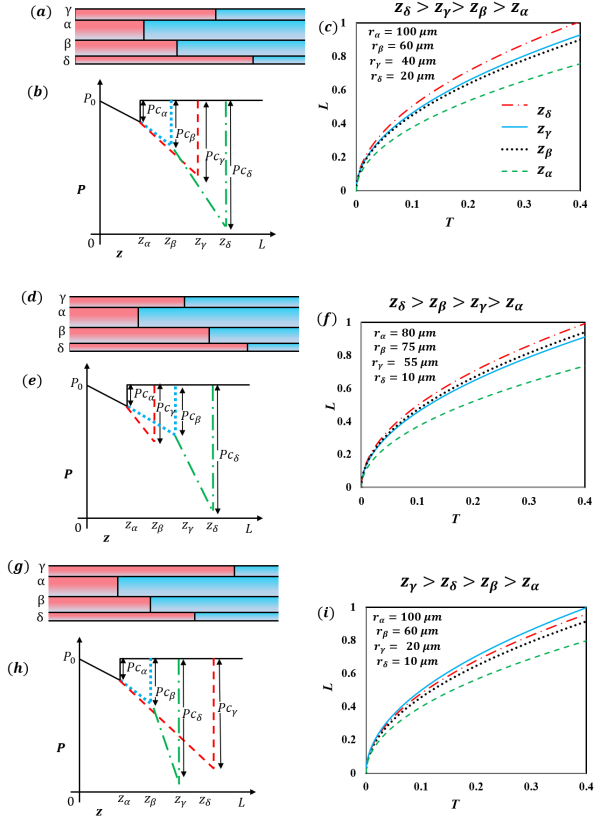}
    \caption{Spontaneous imbibition in interacting four-capillary system where the capillaries are placed in the arrangement $\gamma\alpha\beta\delta$. (a), (d), (g) represent the schematics of all the possible menisci locations, (b), (e), (h) show the corresponding pressure gradient plots. The dimensionless length vs dimensionless time graphs in (c), (f), (i) show radii for which the imbibition behavior shown in (a), (d), (g) is observed. The time at which the invading fluid reaches the outlet end $(T_{bt})$ for the cases (c), (f) and (i) are $0.38$, $0.42$ and $0.38$, respectively.}
    \label{f7}
\end{figure}
We now consider two other random arrangements $\gamma\delta\alpha\beta$ and $\gamma\alpha\beta\delta$, which are shown in Figs. \ref{f6} and \ref{f7}, respectively. In Figs. \ref{f6} and \ref{f7}, we show the schematic of the possible menisci locations during imbibition in (a),(d),(g), while in (b),(e),(h), we show the corresponding pressure gradient plots. We solve the governing equations and observe from the imbibition length vs time plots shown in (c),(f),(i) of Figs. \ref{f6} and \ref{f7} that, the flow behavior anticipated by the pressure gradient graphs by using a suitable contrast in the radii of capillaries. 
\begin{figure}
    \centering
    \includegraphics[width=8 cm]{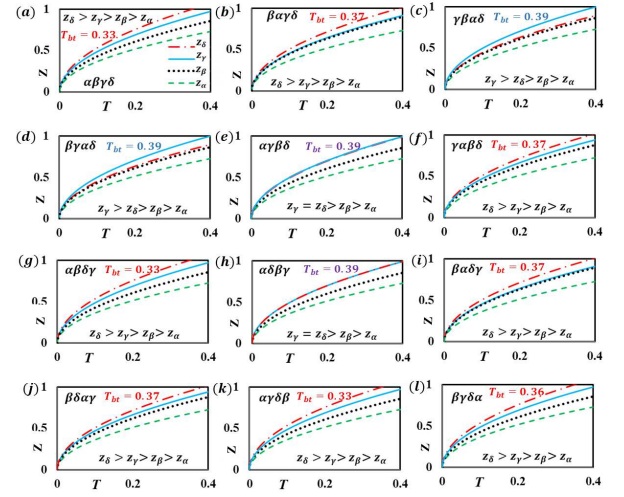}
    \caption{Spontaneous imbibition in interacting four-capillary system. The imbibition behavior for all the $12$ cases of a four-capillary system are represented from (a) to (l) for $r_\alpha=80$ $\mu$m, $r_\beta=60$ $\mu m$, $r_\gamma=40$ $\mu m$ and $r_\delta=20$ $\mu m$. The arrangement and the relative menisci locations for each of the cases (a) to (l) is mentioned in the plots. The breakthrough time $(T_{bt})$ for the cases (a) to (l) are $0.33$, $0.37$, $0.39$, $0.39$, $0.39$, $0.37$, $0.33$, $0.39$, $0.37$, $0.37$, $0.33$, $0.36$.}
    \label{f8}
\end{figure}
From Figs. \ref{f6} and \ref{f7}, we show the effect of the contrast in the capillary radii when the arrangement is same. We see that for the imbibition length vs time shown in Figs. \ref{f5}(f), \ref{f6}(f), \ref{f7}(c), the radii of the capillaries in the interacting capillary system are same although the arrangement of the capillaries is different. We observe that, for the arrangement $\beta\gamma\alpha\delta$ shown in Fig. \ref{f5}(f), the menisci positions are $Z_\gamma>Z_\beta>Z_\delta>Z_\alpha$. For the arrangement $\gamma\delta\alpha\beta$ shown in Fig. \ref{f6}(f), the menisci positions are $Z_\delta>Z_\beta>Z_\gamma>Z_\alpha$ and for the arrangement $\gamma\alpha\beta\delta$ shown in Fig. \ref{f7}(c), the menisci positions are $Z_\delta>Z_\gamma>Z_\beta>Z_\alpha$. Therefore, for an interacting multi-capillary system, the contrast in capillary radii and the arrangement of the capillaries are crucial in determining the imbibition behavior. We can anticipate different meniscus positions based on pressure gradient plots which help in developing governing equations. Fig. \ref{f5}, \ref{f6} and \ref{f7} also report the breakthrough time, $T_{bt}$, i.e., the dimensionless time at which the imbibing fluid reaches the dimensionless length $1$ in one of the interacting capillaries. We also show the capillary radius through which fluid breakthrough occurs. Figs. \ref{f5}(b), \ref{f6}(f) and \ref{f7}(c) indicate that, the breakthrough time and the capillary through which the breakthrough occurs also change with the change in arrangement of capillaries although the contrast in the radii of capillaries remains same. 

We also show the imbibition phenomenon in an interacting four-capillary system for all the twelve possible arrangements as shown in Fig. \ref{f8}. The radii of the capillaries are $r_\alpha=80$ $\mu$m, $r_\beta=60$ $\mu$m, $r_\gamma=40$ $\mu$m and $r_\delta=20$ $\mu$m for all the arrangements. We see that the leading meniscus is in $\delta$ for arrangements shown in Fig. \ref{f8}(a),(b),(f),(g),(i),(j),(k),(l). For the arrangements shown in Fig. \ref{f8}(c),(d), the leading meniscus is in $\gamma$. We see that for arrangements shown in Fig. \ref{f8}(e),(h), the capillaries $\gamma$ and $\delta$ impregnate the same distance with time. We see that, the breakthrough times for all the arrangements are different and vary from $T=0.33$ to $T=0.40$. It can also be observed that, the minimal breakthrough time is in arrangements (a), (g), (k) and (l), shown in Fig. \ref{f8}, which is $0.33$. We have determined  that, for a wetting fluid of viscosity $1$ cp, having a surface tension of $73$ dynes/cm imbibing in the empty capillary system of length $1$ m and having a maximum capillary radius of $200$ $\mu$m, the dimensionless time corresponding to $T=0.01$ is $2.7$ s. So, the breakthrough for the arrangements (a), (g), (k) and (l) shown in Fig. \ref{f8} occurs between $90$ s and $93$ s. For the four capillary system, we can summarize that the arrangement of the capillaries and the capillary radii contrast significantly affects the breakthrough time and the capillary through which the breakthrough occurs.

\subsection{Interacting twenty-capillary system}
\begin{figure}[h!]
    \centering
    \includegraphics[width=8 cm]{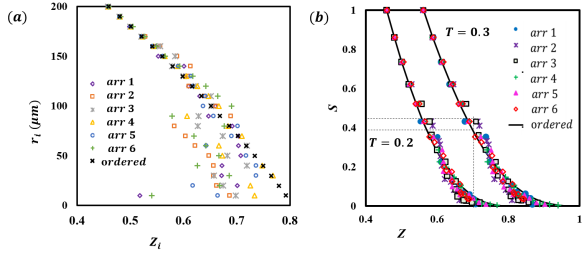}
    \caption{Spontaneous imbibition in an interacting twenty-capillary system in $6$ random arrangements and an ordered arrangement of an interacting twenty-capillary system at $T=0.2$.; (a) radii vs imbibition length,  (b) saturation vs length.}
    \label{f9}
\end{figure}

For an interacting multi-capillary system, the capillary having the leading meniscus and the breakthrough time depends on the contrast in the capillary radii and the arrangement of capillaries in the capillary system. Now, we use the generalized model to predict the imbibition behavior in an interacting capillary system consisting of $n=20$ capillaries and will focus on the arrangement on the capillaries. The number of possible combinations of arrangements for a twenty interacting capillary system is $1.2\times 10^{18}$. The generalized model developed in this study is used to find the menisci location for $1000$ random arrangements of an interacting twenty-capillary system. The radii of the capillaries is a uniform distribution with the maximum radius being $200$ $\mu$m and the minimum radius being $10$ $\mu$m. We show in Fig. \ref{f9}(a), the imbibition length in the capillaries vs the radii of the capillaries at the dimensionless time $T=0.2$ for $6$ random arrangements and the ordered arrangement. We have chosen the $6$ random arrangements such that the disparity in the breakthrough time and the capillary radius through which the breakthrough occurs can be observed for the given radii contrast of the capillaries. We see from Fig. \ref{f9}(a) that, at $T=0.2$, the capillary having the leading meniscus is different for different arrangements and the menisci positions in the capillaries are also dependent on the arrangement. For instance, at $T=0.2$, the meniscus in the small radius capillary of radius $10$ $\mu$m has traveled a dimensionless length of $0.79$ for ordered arrangement, whereas for arrangement $1$, the dimensionless length invaded by the smallest capillary is $0.51$. The saturation at a given imbibition length $Z$ can be defined as the ratio of the cross-sectional area occupied by imbibing fluid to the total cross-sectional area of the interacting capillary system at $Z$, which is $\frac{\sum r_f^2}{\sum_{i=1}^n r_i^2}$, where $r_f$ are the radii of capillaries filled with the imbibing fluid.  We show the saturation vs distance plot in Fig. \ref{f9}(b), at $T=0.2$ and $T=0.3$ for all the $7$ arrangements. The saturation profile of the interacting capillary system changes with a change in the arrangement of the capillaries. For example, the saturation at $Z=0.7$ is $0.43$ for arrangement $3$, and is $0.35$ for the ordered arrangement as indicated by the dashed lines of Fig. \ref{f9}(b).

\begin{figure}[h!]
    \centering
    \includegraphics[width=8 cm]{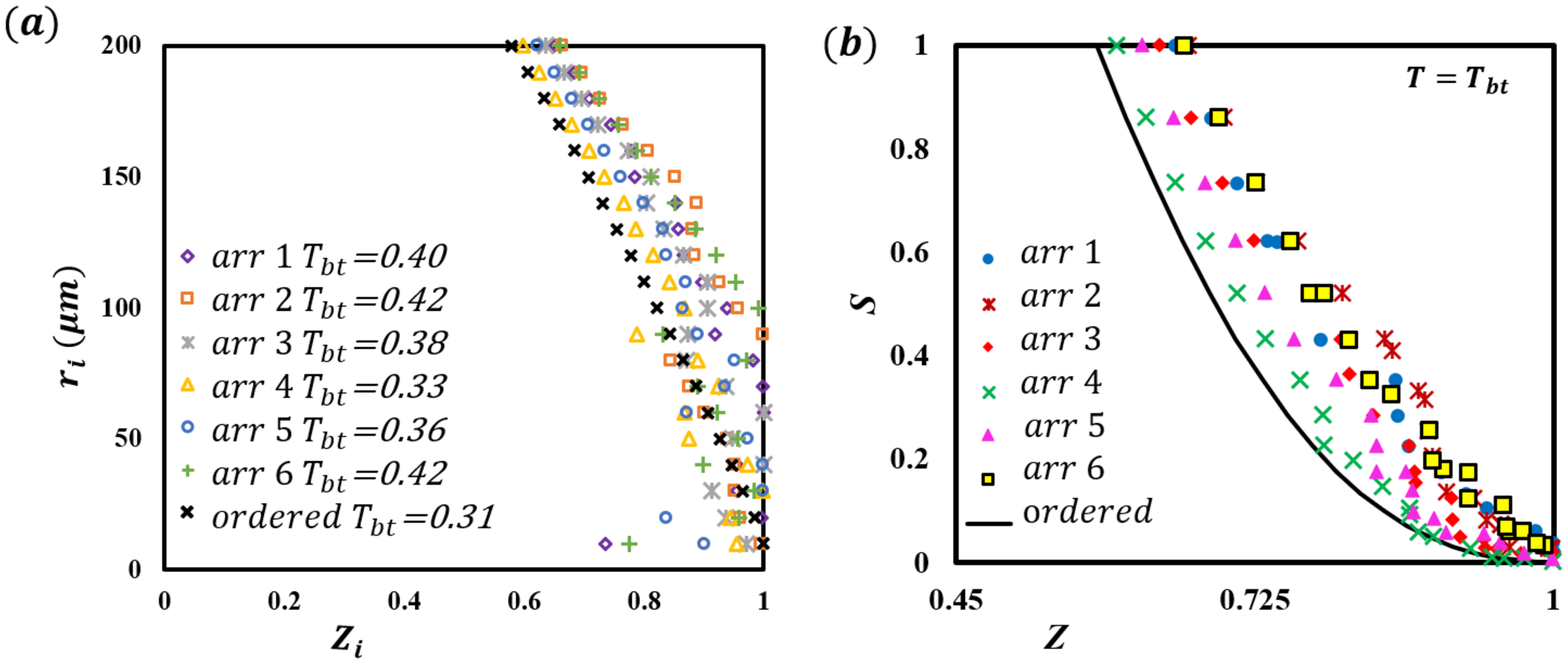}
    \caption{Spontaneous imbibition in an interacting twenty-capillary system in $6$ random arrangements and an ordered arrangement of an interacting twenty-capillary system at breakthrough, $T=T_{bt}$.; (a) radii vs imbibition length,  (b) saturation vs length.}
    \label{f10}
\end{figure}

In Fig. \ref{f10}(a), we show the imbibition in interacting twenty-capillary systems at breakthrough time. We see that, the breakthrough in the interacting twenty-capillary systems occurs through different capillaries at different times for the $6$ random arrangements and the ordered arrangement. The breakthrough time for different arrangements is shown along with the legend of the arrangement in Fig. \ref{f10}(a). In Fig. \ref{f10}(b) we show the saturation vs distance graph at breakthrough time for all the $7$ arrangements. We see from Fig. \ref{f10}(b) that, different amount of the non-wetting fluid is displaced at the time of breakthrough. This shows that the arrangement of the capillaries significantly affects our understanding of the spontaneous imbibition in the porous medium if we use the interacting capillary system as a proxy.

However, if the contrast in the radii of the capillaries is same for different arrangements, therefore, the effective capillary suction causing the imbibition phenomena remains the same. Therefore, at a given time $T$, the total volume imbibed in the interacting capillary system will be the same for all arrangements, which is determined as $\frac{\sum_{i=1}^n r_i^2 Z_i}{\sum_{i=1}^n r_i^2}$. The fraction of the interacting capillary system occupied with the imbibing phase at $T=0.2$ is $0.55$ and at $T=0.3$, the total volume imbibed is $0.67$ for all the $7$ arrangements. However, this is applicable till the breakthrough occurs in one of the arrangements. 

\begin{figure}[h!]
    \centering
    \includegraphics[width=8 cm]{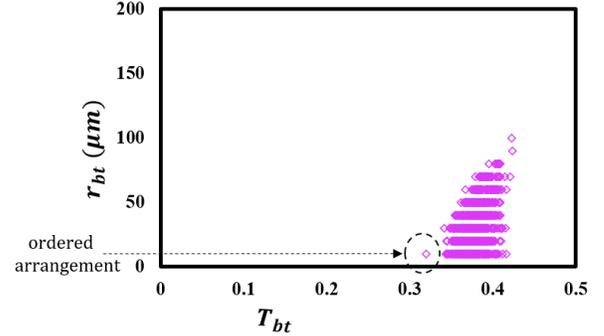}
    \caption{The radii of the capillaries vs breakthrough time in $1000$ arrangements of an interacting twenty-capillary system. It can be observed that, the shortest breakthrough time is in the ordered arrangement, at $T=0.31$ and the maximum breakthrough time observed is at $T=0.42$. The maximum radius of the capillary through which the breakthrough occurs is $100$ $\mu$m while the minimum radius of the capillary through which the breakthrough occurs is $10$ $\mu$m.}
    \label{f11}
\end{figure}

In Fig. \ref{f11}, we show the radius of the capillary having the leading meniscus vs the breakthrough time for the $1000$ randomly chosen arrangements. We see that, the breakthrough for a twenty-capillary system occurs between $T=0.31$ and $T=0.42$, which correspond to dimensional times of $84.93$ s and $115.06$ s, when a wetting fluid of viscosity $1$ cp having a surface tension of $73$ dynes/cm imbibes an interacting capillary system of length $1$ m, having a maximum capillary radius of $200$ $\mu$m. Therefore, for the same contrast in the capillary radii, the maximum and minimum breakthrough time are more than $30$ s apart, indicating that the breakthrough time significantly depends on the arrangement of the capillaries. It can also be observed from Fig. \ref{f11} that, the breakthrough in an ordered multi-capillary system occurs through the smallest radius capillary at $T=0.31$, which is the least breakthrough time as compared to other arrangements. Fig. \ref{f11} also shows that, the maximum radius of the capillary through which the breakthrough occurs is as large as $100$ $\mu$m, while the minimum radius of the capillary through which the breakthrough occurs is $10$ $\mu$m. We see from Fig. \ref{f10}(a) that, for arrangement $6$ indicated by a $+$ sign in the legend, the leading meniscus is in the $100$ $\mu$m radius capillary through which the breakthrough occurs at $T_{bt}=0.42$. From Fig. \ref{f11}, we also see that, when the breakthrough occurs through the smallest radius capillary, the breakthrough time may vary from $T=0.31$ to $T=0.41$, for which the total volume fraction of the interacting capillary system occupied by the invading phase can lie between $0.69$ and $0.79$. For instance, if the breakthrough occurs through the $70$ $\mu$m radius capillary, the breakthrough time lies between $T=0.38$ and $T=0.42$ and the total volume fraction imbibed by the wetting phase lies between $0.76$ and $0.8$.

\subsubsection{Comparison of bundle-of-tubes model and interacting multi-capillary system}

\begin{figure}[h!]
    \centering
    \includegraphics[width=8 cm]{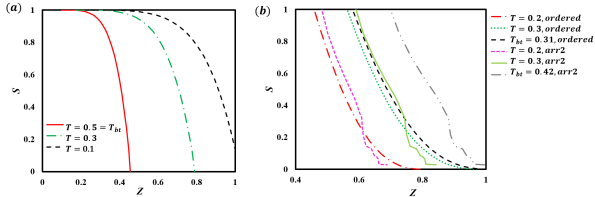}
    \caption{(a) Saturation vs length during spontaneous imbibition in bundle-of-tubes consisting of twenty non-interacting capillaries at $T=0.1$, $T=0.3$, and $T=T_{bt}=0.5$. (a) Saturation vs length during spontaneous imbibition in interacting twenty-capillary system for ordered arrangement and arrangement $2$ at $T=0.1$, $T=0.3$, $T=T_{bt}$.}
    \label{f12}
\end{figure}

Comparing the bundle-of-tubes and the interacting capillary models, we show the saturation across the cross-section vs length imbibed for bundle-of-tubes model in Fig \ref{f12}(a) and for two arrangements of the interacting twenty-capillary system in Fig \ref{f12}(b). In Fig. \ref{f12}(a), we show the saturation across the cross-section vs length at $T=0.1$, $T=0.3$ and $T=T_{bt}=0.5$ for the non-interacting capillary system having $n=20$ capillaries, where the capillary radii are same as the capillary radii considered in the interacting multi-capillary system. We know that, in a bundle-of-tubes model, the imbibition follows Washburn's diffusive dynamics and therefore the invaded length is the longest in the largest radius capillary as compared to other capillaries. 

For non-interacting capillaries, by non dimensionalizing Eq. \ref{eq02}, we obtain,
\begin{equation}
    \label{eq24}
    Z_i^2 = 2\epsilon_i \lambda_i^2 T,
\end{equation}
where, $Z_i=\frac{z_i}{L}$ is the dimensionless length imbibed in a capillary having radius $r_i$ and $L$ is the total length of the capillary system. The time is non-dimensionalised as $T=\frac{Pc_\alpha r_\alpha^2}{8\mu L^2}t$. In Eq. \ref{eq24}, $\epsilon = \frac{Pc_i}{Pc_\alpha}$ and $\lambda_i = \frac{r_i}{r_\alpha}$. Here $Pc_\alpha$ and $r_\alpha$ are the capillary pressure and radius of the largest radius capillary which has $200$ $\mu$m radius. The maximum value of $\epsilon_i$ and $\lambda_i$ are $1$, which is for the largest radius capillary. For all other capillaries $\epsilon_i$ and $\lambda_i$ are less than $1$. 

We see from Fig. \ref{f12}(a) that, due to the large cross-section area of the large radius capillary, the impregnated fluid volume in it contributes to a large fraction of cross-sectional saturation for the bundle-of tubes model. The breakthrough in bundle-of-tubes model always occurs through the largest capillary unlike the interacting capillary system where the largest radius capillary always has a lagging meniscus. Therefore, the time taken for the bundle-of-tubes model for breakthrough is $136.98$ s, at $T=0.5$, at which the total volume of imbibition is $0.86$. The fractional volume occupied at breakthrough in a bundle-of-tubes model is considerably large as compared to the fractional volumes occupied in the interacting capillary system for the same contrast in radii, which lies between $0.69$ and $0.79$.
In Fig. \ref{f12}(b), we show the saturation across the cross-section vs length at $T=0.1$, $T=0.3$, and $T=T_{bt}$ for an ordered interacting capillary system and for the arrangement $2$ shown in Figs. \ref{f9}, \ref{f10}. The leading meniscus for an orderly arranged interacting capillary system is in the smallest radius capillary and we know that the fraction of saturation contributed by the smallest radius capillary is little. For the arrangement $2$ shown in Fig. \ref{f12}(b), the leading meniscus is in the capillary having a radius of $90$ $\mu$m. We also evaluated that, for the bundle of tubes model, the cross-section area of the leading capillary is $13\%$ of the total cross-section area, whereas for the ordered arrangement and the arrangement $2$, the respective cross-section area of the leading meniscus capillaries are $0.03\%$ and $2.82\%$. Therefore, it can be observed from Fig. \ref{f12} that, the cross-section saturation gradually decreases for the bundle-of-tubes model while for the interacting capillaries, there is a steep decrease in the cross-sectional saturation with length. 
We also see that, the leading meniscus in the largest radius capillary for the bundle-of-tubes model reaches the exit end at $T=0.5$ as shown in Fig. \ref{f12}(a), whereas for the orderly arranged interacting capillary system, and arrangement $2$, the breakthrough times are $T=0.31$ and $T=0.42$, respectively, as shown in Fig. \ref{f12}(b), which correspond to dimensional times of $84.93$ s and $115.06$ s. Therefore, the saturation of the porous medium with length and the breakthrough time significantly differ for the bundle-of-tubes model and for the different arrangements of the interacting multi-capillary system, although the contrast in the radii of the capillaries is the same. In real porous media, the imbibing fluid saturation decreases gradually with length, similar to the trend shown by the interacting multi-capillary system. 

The saturation vs the length curves from the interacting multi-capillary system are consistent with the observations from the imbibition phenomena in porous media described by Dong et al., Ding et al., Debbabi et al., and Akbari et al.,\cite{dong1998characterization,ding2020application,debbabi2017viscous,akbari2019new}. Further, it was also experimentally shown by Bico and Qu{\'e}r{\'e} that, the propagating fluid front in a porous medium has two fronts; a leading microscopic front and a lagging macroscopic front\cite{bico2003precursors}. It was also described that the interaction among the pores causes the menisci in the smaller radii pores to lead as compared to the larger radii pores\cite{bico2003precursors}. The current study elaborates this flow behaviour by considering the interaction among the pores and rather than classifying the pores into categories of smaller radii and larger radii, the model developed in this study accurately describes the physics of flow in each of the interacting pores. It is evident from Figs. \ref{f9}(a), \ref{f10}(a) and \ref{f12}(b) that the menisci in most of the capillaries having radii between $10$ $\mu$m and $100$ $\mu$m are ahead of the menisci in the capillaries having larger radii, i.e., the capillaries having radii between $100$ $\mu$m to $200$ $\mu$m. It was also previously described that the lagging macroscopic front is responsible for the saturation of a porous medium\cite{bico2003precursors}, which is in good agreement with the saturation profile anticipated by the interacting multi-capillary system, as shown in Fig. \ref{f12}(b). The saturation profile in the Fig. \ref{f12}(a) for the non-interacting bundle of capillaries shows that the large pores are responsible for the leading macroscopic front and the saturation of the porous medium, which is contrary to the experimental observations in real porous media.\cite{bico2003precursors,Ashraf2017,ashraf2019,ashraf2019generalized,ashraf2019capillary}. 

Although the interacting multi-capillary system can accurately describe the physics of flow in porous systems, the model in its current state has certain limitations. The tortuosity of the interacting pores and the varying pore geometry needs to be taken into account to represent a real porous medium. Further, the model developed in this study predicts the imbibition phenomena until the leading meniscus reaches the exit end of the porous medium. Upon breakthrough, the invasion dynamics change considerably and suitable modifications are needed to be included in the model to predict the post-breakthrough flow dynamics.

\section{Conclusions}
In conclusion, we investigated the imbibition phenomenon of a wetting fluid in a randomly arranged interacting capillary system. We developed a strategy to formulate a one-dimensional lubrication approximation model to predict the flow behavior in an interacting multi-capillary system. The generalized model can predict the imbibition behavior for all the $\frac{n!}{2}$ possible arrangements of an interacting $n$-capillary system. It was observed that, for the same contrast in the radii of the capillaries, the imbibition phenomenon depends significantly on the arrangement of the capillaries within the capillary system. It was also observed that, the imbibition behavior is affected by the contrast in the radii of the capillaries when the arrangement of the capillaries is the same. Further, it was shown that, the arrangement and the contrast in the radii of the capillaries determine the relative menisci location, the capillary having the leading meniscus and the breakthrough time. The cross-sectional saturation of the impregnating fluid along the length of the capillary system also changes with the change in the arrangement of the capillaries. However, the total capillary pressure driving the flow is same for all arrangements, therefore, the overall volume fraction occupied by the invading fluid at a given time remains the same across all arrangements, till breakthrough occurs in one of the arrangements. In the current study, we have shown the radius of the capillary having the leading meniscus for $1000$ different arrangements of an interacting twenty-capillary system having an uniform distribution of radii with the smallest and the largest radii of the capillaries being $10$ $\mu$m and $200$ $\mu$m. We observed that, depending on the arrangement of the capillaries, the leading meniscus can be in any of the capillaries having radii between $10$ $\mu$m and $100$ $\mu$m and the breakthrough time lies between $T_{bt}=0.31$ and $T_{bt}=0.42$. 

We also compared the invasion in an interacting multi-capillary system investigated in this study with the bundle-of-tubes model where the capillaries do not interact with each other. The saturation with length in the bundle-of-tubes model and the ordered interacting multi-capillary system are significantly different from one another. For the bundle-of-tubes model, the leading meniscus is always in the largest radius capillary. For an ordered interacting multi-capillary system, the leading meniscus is always in the smallest radius capillary having a radius of $10$ $\mu$m. It was observed that, the breakthrough in the bundle-of-tubes model occurs at $T_{bt}=0.5=136.98$ s, while the breakthrough in an ordered interacting capillary system occurs at $T_{bt}=0.31=84.93$ s and for one of the randomly arranged interacting capillary system is at $T_{bt}=0.42=115.06$ s. The fractional volume occupied at breakthrough for bundle-of-tubes model is $0.86$, whereas for the interacting capillary system, it is between $0.69$ and $0.79$, respectively. The generalized model developed in this study is useful in understanding pore scale behavior during spontaneous imbibition applications in developing design based porous media like loop heat pipes, diagnostic devices and microfluidic devices; or predicting the flow behaviour in real porous medium such as oil reservoirs. 
\section*{Conflicts of interest}
There are no conflicts to declare.


\section*{Acknowledgement}
The authors acknowledge the financial support granted by Science and Engineering Board (SERB), India (DiaryNo. SERB/F/1297/2017-1, File No. ECR/2017/000257).




\bibliography{PoF.bib}
\end{document}